\documentclass[reprint, superscriptaddress, longbibliography, floatfix, twocolumn, nofootinbib, amsmath, amssymb,
 aps, showkeys]{revtex4-2}
\usepackage{mathtools}
\usepackage{xfrac}
\usepackage{braket}
\usepackage{siunitx}
\usepackage{hyperref}
\usepackage{gensymb}
\usepackage{xcolor}
\usepackage{graphicx}
\usepackage{dcolumn}
\usepackage{bm}
\usepackage{dsfont}
\usepackage{booktabs}
\usepackage{soul}

\usepackage{overpic}

\makeatletter
\newcommand*{\rom}[1]{\expandafter\@slowromancap\romannumeral #1@}
\makeatother

\newcommand{\ra}[1]{\renewcommand{\arraystretch}{#1}} 

\begin{document}

\preprint{APS/123-QED}

\title{Indistinguishability of single Raman photons from single atoms}

\author{Pascal Baumgart}
\affiliation{Fachrichtung Physik, Universit\"at des Saarlandes, 66123 Saarbr\"ucken, Germany}
\affiliation{Zentrum für Quantentechnologien (QuTe), Universit\"at des Saarlandes, 66123 Saarbr\"ucken, Germany}

\author{Max Bergerhoff}
\affiliation{Fachrichtung Physik, Universit\"at des Saarlandes, 66123 Saarbr\"ucken, Germany}
\affiliation{Zentrum für Quantentechnologien (QuTe), Universit\"at des Saarlandes, 66123 Saarbr\"ucken, Germany}

\author{J\"urgen Eschner}
\email{juergen.eschner@physik.uni-saarland.de}
\affiliation{Fachrichtung Physik, Universit\"at des Saarlandes, 66123 Saarbr\"ucken, Germany}
\affiliation{Zentrum für Quantentechnologien (QuTe), Universit\"at des Saarlandes, 66123 Saarbr\"ucken, Germany}

\date{\today}

\begin{abstract}	
We theoretically investigate the indistinguishability of single photons generated from single trapped $^{40}$Ca$^+$ ions in a Raman scattering process driven by few-nanosecond excitation pulses. Of particular interest is how spontaneous decay back to the initial state affects Hong-Ou-Mandel (HOM) photon interference. Numerical simulations identify the mean number of back-decays as a measurable quantity that correlates with achievable HOM visibility. Optimization of the excitation pulse with respect to a trade-off between photon yield and indistinguishability is analyzed. Finally, we compare the performance of different trapped-ion species for long-range dual-rail entanglement swapping via HOM photon interference. 
\end{abstract}


\maketitle

The interconnection of macroscopically separated quantum memories via 
photon-mediated entanglement \cite{Simon_2003, Moehring_2007} is one of the pivotal resources for scaling up the performance of various quantum technologies. The use of quantum repeater schemes \cite{Briegel_1998, vanLoock_2020} together with quantum frequency conversion 
 \cite{Bock_2018, Saha_2023} enables the distribution of entanglement over large distances, paving the way towards large-scale quantum networks \cite{Kimble_2008, Wehner_2018} and their applications in secure quantum communication \cite{Schwonnek_2021, Nadlinger_2022}, distributed quantum computing \cite{Cirac_1999, Jiang_2007, Monroe_2014}, and quantum sensing \cite{Zhang_2021, Malia_2022, Zhao_2021}. On shorter scales and without quantum frequency conversion, the use of entanglement swapping 
protocols based on photonic Bell state measurements can efficiently connect quantum processing units in a modular quantum computation architecture \cite{Main_2025, OReilly_2024}. 

These applications are facilitated by the natural and stable source of entanglement that single-photon emission provides \cite{Blinov_2004, Bock_2018}. For performing photonic Bell state measurements, a crucial additional requirement is the ability to create single photons capable of high-contrast Hong-Ou-Mandel (HOM) interference \cite{Hong_1987}, i.e., photons with a high degree of indistinguishability. 

Many species of trapped atoms or ions 
offer the possibility of generating single photons via spontaneous Raman scattering involving a stable ground state, a short-lived excited state, and a third metastable state. 
During excitation, however, the emitter might spontaneously decay back to the initially prepared state and be re-excited---possibly multiple times---before the final photon 
is emitted. These back-decay events must be taken into account when 
determining the spectral and temporal properties of the 
Raman photon. In earlier work \cite{Mueller_2017}, it was shown that 
the spectrum is not affected by these back-decays.  
The temporal profile, however, is altered, since each additional back-decay adds a time-shifted contribution to the photon's full temporal 
wave packet, resulting in a photon that is no longer Fourier-limited. 
This degrades the indistinguishability of two identical copies 
of the photon, resulting in reduced HOM interference visibility. 

Back-decays to the initial state during excitation may be avoided by using excitation pulses with pulse lengths shorter than (or at most comparable to) the excited-state lifetime, which is typically on the order of a few nanoseconds. 
Good interference properties have been demonstrated with photons generated with excitation pulses that are much shorter than the excited-state lifetime \cite{Maunz_2007, Kim_2020}. But since the laser power required for maximal population transfer to the excited state scales with the inverse square of the pulse length, it might be desirable to use excitation pulses with pulse lengths closer to the lifetime of the excited state. In some cases, the use of sub-nanosecond pulses may even have to be avoided entirely since they cause unwanted excitation to nearby hyperfine states \cite{vanLeent_2022}. 

Here, we investigate the feasibility of using few-nanosecond excitation pulses for generating indistinguishable photons via spontaneous Raman scattering. With numerical calculations based on quantum trajectories \cite{Plenio_1998} and the quantum regression theorem for time-dependent Hamiltonians \cite{Fischer_2016}, we extract statistical measures related to the aforementioned back-decays---which become significant in this regime---and show for various pulse lengths and strengths how these quantities correlate with achievable HOM visibility. The calculations are applied to the generation of 854-nm photons from single trapped $^{40}$Ca$^+$ ions, a relevant case in our quantum communication experiments \cite{Bock_2018, Bergerhoff_2024, Bergerhoff_2026, Haen_2026}. The results expand and generalize the experimental findings in \cite{Baumgart_2026exp}, which identified the mean number of back-decays as an easily accessible single-emitter quantity that correlates with interference properties in an experiment involving two emitters. Here, a larger parameter region for the excitation pulses is probed numerically, and the optimization of these parameters with respect to photon yield and indistinguishability is investigated. 

The present model is applicable to general single-photon generation schemes involving a Raman transition to a (meta-)stable final state, and thus to a variety of single photon sources, including different trapped-ion and atom species. In a final section of the paper, we evaluate the model for trapped-ion species commonly used in quantum networking applications and compare their performance for long-range dual-rail entanglement swapping.

\section{Model}\label{sec:model}

\begin{figure}[h]
    \centering
        \begin{overpic}[width=\linewidth]{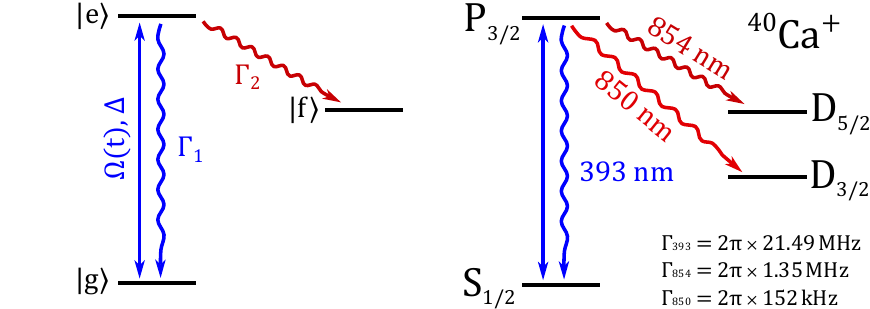}
	       	\put (2,33) {(a)}
            \put (47,33) {(b)}
        \end{overpic}
    \caption{(a) Three-level system in the $\Lambda$-configuration with a stable ground state $\ket{g}$, short-lived excited state $\ket{e}$, and a (meta-)stable final state $\ket{f}$. The $\ket{g}\leftrightarrow\ket{e}$ transition is driven by a laser pulse with time-dependent Rabi frequency $\Omega(t)$ and detuning $\Delta$. The excited state decays to the two (meta-)stable states with spontaneous decay constants $\Gamma_1$ and $\Gamma_2$. (b) Relevant level scheme of $^{40}$Ca$^+$ with transition wavelengths and spontaneous decay constants.}
    \label{fig:threelevelscheme}
\end{figure}

A three-level system in the $\Lambda$-configuration is considered, with a stable ground state $\ket{g}$, a short-lived excited state $\ket{e}$, and a (meta-)stable final state $\ket{f}$, as illustrated in \autoref{fig:threelevelscheme} (a). The ground-to-excited-state transition is driven by a laser pulse, modeled by a time-dependent Rabi frequency $\Omega(t)$, which is detuned from resonance by $\Delta$. Photons are spontaneously emitted on the transitions $\ket{e}\to\ket{g}$ and $\ket{e}\to\ket{f}$ with spontaneous decay constants $\Gamma_1$ and $\Gamma_2$, respectively. 
\vspace{1mm}

The system is modeled with a master equation in Lindblad form 
\begin{align}
	&\frac{\partial}{\partial t}\rho(t)=\mathcal{L}(t)\rho(t)\nonumber\\
	&=-\frac{i}{\hbar}\left[H(t),\rho\right]+\sum_{k\in\{1,2\}} 
	C_k\rho(t)C_k^\dagger-\frac{1}{2}\left\{C_kC_k^\dagger,\rho(t)\right\}\label{eq:masterequation}
\end{align} 
with the coherent Hamiltonian 
\begin{align}
	H(t)=H_0+f(t)\cdot H_{\mathrm{I}}\label{eq:Ht}
\end{align} 
describing the ion interacting with a laser whose Rabi frequency varies in time according to the temporal profile 
$f(t)$, such that $\Omega(t)=\Omega_0f(t)$ with the constant Rabi frequency $\Omega_0$. The collapse operators $C_k$ correspond to the two spontaneous decay paths with spontaneous decay constants $\Gamma_k$. The total decay rate of the excited state is $\Gamma=\Gamma_1+\Gamma_2$.
The operators according to \autoref{eq:masterequation} and \autoref{eq:Ht} are given by 
\begin{align}
	H_0&=\hbar\Delta\vert e\rangle\langle e\vert,~H_{\mathrm{I}}=f(t)\cdot \frac{\hbar\Omega_0}{2}\left(
	\vert e\rangle\langle g\vert + \vert g\rangle\langle e\vert
	\right),\\
	C_1 &=  \sqrt{\Gamma_1}\vert g\rangle\langle e\vert,~C_2=\sqrt{\Gamma_2}\vert f\rangle\langle e\vert,
\end{align}
in the reference frame rotating with the laser frequency, and 
assuming a real-valued Rabi frequency $\Omega_0$. 

In \autoref{sec:simresults}, the model will be applied to the $\mathrm{S}_{1/2}\to\mathrm{P}_{3/2}\to\mathrm{D}_{5/2}$ Raman transition in a single trapped $^{40}$Ca$^+$ ion, which is excited with a laser pulse at 393\,nm and emits a single photon at 854\,nm. There is also a parasitic decay channel from $\mathrm{P}_{3/2}$ to $\mathrm{D}_{3/2}$ under emission of an 850-nm photon. The relevant level structure, including the transition wavelengths and spontaneous decay constants, is shown in \autoref{fig:threelevelscheme} (b).
In this configuration, the ground state corresponds to the $\mathrm{S}_{1/2}$ manifold, the excited state to $\mathrm{P}_{3/2}$, and in the final state, $\mathrm{D}_{3/2}$ and $\mathrm{D}_{5/2}$ are combined. The spontaneous decay constants are then given by $\Gamma_1 = \Gamma_{393} = 2\pi \times 21.49$\,MHz and $\Gamma_2 = \Gamma_{854} + \Gamma_{850} = 2\pi \times 1.502$\,MHz \cite{Gerritsma_2008}.
Correspondingly, the success probability for creating a Raman photon, $P_{\mathrm{Raman}}$, also combines spontaneous emission into the two final states. The probability $P_{854}$ of creating a Raman photon specifically at 854\,nm is then found as $P_{854} = \Gamma_{854}/(\Gamma_{854}+\Gamma_{850})\cdot P_{\mathrm{Raman}}$. The reduction to an effective three-level system 
is valid for most of the relevant experimental systems, as detailed in appendix \ref{appendix:threelevel}.

\subsection{Back-decay statistics: quantum jump decomposition}\label{sec:quantumjumps}

To investigate the effect of back-decays on the interference properties of a single Raman photon, a figure of merit has to be introduced to quantify the statistics of these events. The full information about the number distribution of back-decays is contained in the conditional probabilities $P_N$ ($N\geq 0$) that exactly $N$ photons were scattered via a back-decay to the ground state during the excitation pulse, given that the final photon was a Raman photon. 
Especially the probability $P_0$ for the Raman photon to be emitted without any prior back-decays may be considered as a figure of merit, with $P_0=1$ indicating a Fourier-limited photon \cite{Mueller_2017}. 

Another figure of merit is the mean number of back-decays 
\begin{align}
    \langle N\rangle=\sum_{N=0}^\infty N\cdot P_N\label{eq:npn}
\end{align}
 before the emission of a Raman photon, with $\langle N\rangle=0$, equivalent to $P_0=1$, being the optimal value with respect to temporal indistinguishability. 
As demonstrated in \cite{Baumgart_2026exp}, this quantity is also experimentally easily accessible. 

Finally, also the success probability $P_{\mathrm{Raman}}$ that a Raman photon was emitted after a single excitation pulse is of great practical interest. 

In order to obtain access to the desired photon statistics from the atomic density matrix, its quantum jump decomposition
\begin{align}
	\rho(t)=\sum_{n=0}^\infty\sum_{m=0}^1 \rho^{(n,m)}(t)
\end{align}
is investigated, where $\rho^{(n,m)}(t)$ corresponds to the contribution resulting in exactly $n$ photons emitted via a decay to the ground state, and $m$ photons via a decay to the final state. 
By setting $m$ maximally to 1, it is already considered that at most a single photon is scattered on the $\ket{e}\to\ket{f}$ transition, and that this decay terminates the evolution of the system. 

Following the description in \cite{Plenio_1998}, the quantum jump decomposition is obtained by splitting up the master equation
\begin{align}
	\frac{\partial}{\partial t}\rho(t)=\mathcal{L}(t)\rho(t)=\mathcal{L}_{\mathrm{eff}}(t)\rho(t)+\sum_{k}\mathcal{J}_k\rho(t)
\end{align}
into the time-evolution with the non-Hermitian effective Hamiltonian 
\begin{align}
	\mathcal{L}_{\mathrm{eff}}(t)\rho(t)&=-\frac{i}{\hbar}\left(H_{\mathrm{eff}}(t)\rho(t)-\rho(t)H^\dagger_{\mathrm{eff}}(t)\right), \\
	H_{\mathrm{eff}}(t)&=H(t)-\frac{i\hbar}{2}\sum_kC_k^\dagger C_k
\end{align}
and the application of the jump-operators 
\begin{align}
	\mathcal{J}_k\rho(t)=C_k\rho(t)C_k^\dagger
\end{align}
for $k=1$ and $k=2$. Given an initial state $\rho(t_0)$ at time $t_0$, the zero-photon contribution to the density 
matrix is then given by 
\begin{align}
	\rho^{(0,0)}(t)&=U(t,t_0)\rho(t_0)\nonumber\\
    &=\mathcal{T}\exp\left(\int_{t_0}^{t}\mathrm{d}t^\prime\,\mathcal{L}_{\mathrm{eff}}(t^\prime)\right)\rho(t_0), 
\end{align} 
the formal solution to $\partial_t\rho(t)=\mathcal{L}_{\mathrm{eff}}(t)\rho(t)$, where $\mathcal{T}$ is the time-ordering operator. 
For a time-independent drive $f(t)=1$, the propagator $U(t,t_0)$ explicitly evaluates to 
\begin{align}
	U(t,t_0)\rho(t_0)=e^{-\frac{i}{\hbar}H_{\mathrm{eff}}(t-t_0)}\rho(t_0)e^{\frac{i}{\hbar}H_{\mathrm{eff}}(t-t_0)}.  
\end{align}
A varying temporal profile $f(t)$ may be taken care of by approximating it as stepwise time-independent \cite{Campaioli_2024}. The other contributions 
to the quantum jump decomposition are then iteratively computed via the relations  
\begin{align}
	\rho^{(n+1,m)}(t)&=\int\limits_{t_0}^t\mathrm{d}t^\prime\, U(t,t^\prime)\mathcal{J}_1\rho^{(n,m)}(t^\prime), \\
	\rho^{(n,m+1)}(t)&=\int\limits_{t_0}^t\mathrm{d}t^\prime\, U(t,t^\prime)\mathcal{J}_2\rho^{(n,m)}(t^\prime).
\end{align}
From these contributions to the full density matrix $\rho(t)$, one extracts the time-dependent probabilities 
\begin{align}
    P^{(n,m)}(t)=\mathrm{Tr}(\rho^{(n,m)}(t))\label{eq:pnm}
\end{align}
that exactly $n$ photons have been emitted via a back-decay to the ground state and $m$ photons via a decay to the final state. 

In order to obtain the above-mentioned figures of merit, one starts by considering the dynamics conditioned on the emission of a Raman photon, i.e., the sum of all quantum jump contributions to the density matrix that result in such a photon 
\begin{align}
    \rho_1(t)&=\sum_{n=0}^\infty \rho^{(n,1)}(t)\nonumber\\
    &=\int_{t_0}^{t}\mathrm{d}t^\prime\,U(t,t^\prime)\mathcal{J}_2\left[\sum_{n=0}^\infty \rho^{(n,0)}(t^\prime)\right]\nonumber\\
    &=\int_{t_0}^{t}\mathrm{d}t^\prime\,U(t,t^\prime)\mathcal{J}_2\rho_0(t^\prime), 
\end{align}
where $\rho_m(t)$ is the sum of all contributions corresponding to the emission of $m$ Raman photons for $m=0$ or $m=1$. By following the derivation of the full quantum jump decomposition \cite{Lucas_2014} while leaving out the $k=2$ jump term, one sees that $\rho_0(t)$ is given by the 
solution to the effective master equation 
\begin{align}
    \frac{\partial}{\partial t}\rho_0(t)=\mathcal{L}_0(t)\rho(t)=\mathcal{L}_{\mathrm{eff}}\rho_0(t)+\mathcal{J}_1\rho_0(t)
\end{align}
which only includes the $\ket{e}\to\ket{g}$ jump term $\mathcal{J}_1$. The success probability for the generation of a Raman photon is now obtained by taking the trace of the conditional dynamics and evaluating it as 
$t\to\infty$: 
\begin{align}
    P_{\mathrm{Raman}}=\lim_{t\to\infty}\mathrm{Tr}(\rho_1(t)),\label{eq:praman}
\end{align}
i.e. the trace is evaluated 
at some time sufficiently larger than both the excitation pulse length and the spontaneous decay time of the excited state. 

The probability $P_N$ for exactly $N$ back-decays conditioned on the emission of a Raman photon is then 
\begin{align}
    P_N&\equiv P(n=N~|~m=1)=\frac{P(n=N~\cap~m=1)}{P(m=1)}\nonumber\\
    &=\frac{\lim_{t\to\infty} P^{(N,1)}(t)}{P_{\mathrm{Raman}}}, 
    \label{eq:PN}
\end{align}
where $P^{(N,1)}(t)$ is defined in \autoref{eq:pnm}. The probabilities $P_N$ give valuable insight into the temporal structure of the Raman photon and how the mean number of back-decays $\langle N\rangle$ arises for different excitation pulse parameters. 

Using \autoref{eq:PN} in \autoref{eq:npn}, the value of $\langle N\rangle$ can be approximated, but as this method converges badly, we employ a more robust way. It considers
the number distribution of back-decay photons with the creation operator $a_1^\dagger(t)$ that precede the emission of a Raman photon with the creation operator $a_2^\dagger(t)$ by some time delay $\tau$. This is expressed by the normally ordered 
two-time second-order correlation function 
\begin{align}
    G^{(2)}_{\mathrm{back-decay}}(t,\tau)=\langle a_1^\dagger(t-\tau)a_2^\dagger(t)a_2(t)a_1(t-\tau)\rangle.\label{eq:g2bd}
\end{align}
In order to compute it, the photonic creation and annihilation operators are related to their corresponding atomic collapse operators $a_k\mapsto C_k$ for $k=1$ and $k=2$, as introduced in \autoref{eq:masterequation} \cite{Gardiner_1985}.  
Then the correlation function is evaluated from the atomic dynamics via a version of the quantum regression theorem for time-dependent Hamiltonians \cite{Fischer_2016}. 
The mean number of back-decays is finally obtained by integrating over the 
emission time $t$ of the Raman photon and the time delay $\tau$, and dividing the result by the success probability $P_{\mathrm{Raman}}$ to generate the Raman photon: 
\begin{align}
    \langle N\rangle = \frac{\Gamma_1\Gamma_2\int_{0}^\infty\mathrm{d}t\int_0^\infty\mathrm{d}\tau~G^{(2)}_{\mathrm{back-decay}}(t,\tau)}{P_{\mathrm{Raman}}}.
\end{align}
The success probability can be calculated as in \autoref{eq:praman}, but 
there are also other ways, for example, by evaluating the 
population in the final state as $t\to\infty$, or by integrating 
over the expectation value of the photon number operator 
$\langle n_2(t)\rangle=\langle a_2^\dagger(t) a_2(t)\rangle$.

\subsection{Hong-Ou-Mandel visibility}\label{sec:hom}

The indistinguishability of two photons 
is typically quantified by the HOM interference visibility. For this, one considers the two photons entering a 50:50 beam splitter through the input modes $a$ and $b$ with corresponding bosonic creation operators $a^\dagger(t)$ and $b^\dagger(t)$. 
Perfectly indistinguishable photons coalesce at the beam splitter such that no coincidences are measured at the two outputs. Varying between parallel and orthogonal mode polarizations is then a convenient way to adjust the indistinguishability between its maximal and minimal value, and the HOM visibility is evaluated as 
\begin{align}
    V=1-\frac{C_\parallel}{C_\perp}, 
\end{align}
where $C_\parallel$ and $C_\perp$ are coincidence counts of photons that enter with parallel and orthogonal polarizations, respectively. With photonic wave packets, the visibility becomes a function $V(T)$ of the  size of the time window, $T$, within which detections at the two outputs are treated as coincident. 

The HOM visibility is computed via the normally ordered two-time second-order correlation function between the output fields of the beam splitter \cite{Ou_1988, Woolley_2013}. 
As we aim to characterize the indistinguishability of photons from one and the same generation process, 
we assume the same time-dependence for the creation operators $a^\dagger(t)$ and $b^\dagger(t)$ at the two inputs, such that their correlation functions are equal. Thus, all correlation functions of $b$ may be replaced with the corresponding ones of $a$, and the full expression, still including a possible angle $\phi$ between the polarizations, takes the form
\begin{align}
    G^{(2)}_{\mathrm{HOM}}&(t,\tau,\phi)=\frac{1}{2} \bigl(
    \langle n_a(t)\rangle\langle n_a(t+\tau)\rangle \nonumber \\
   &-\cos^2\phi~\vert G_a^{(1)}(t,\tau))\vert^2 \bigr). \label{eq:G2hom}
\end{align}
Here $n_a(t)=a^\dagger(t)a(t)$ is the photon number operator, and
$G^{(1)}_a(t,\tau)$ is the first-order optical coherence of the input field 
\cite{Woolley_2013}. 
For a more detailed derivation, see appendix \ref{appendix:details}.

Taking into account the window within which two detections at the outputs are considered coincident, the HOM visibility is given by
\begin{align}
    V(T)=1-\frac{\int_0^\infty\mathrm{d}t\int_{-T/2}^{T/2}\mathrm{d}\tau~G^{(2)}_{\mathrm{HOM}}(t,\tau,0)}{\int_0^\infty\mathrm{d}t\int_{-T/2}^{T/2}\mathrm{d}\tau~G^{(2)}_{\mathrm{HOM}}(t,\tau,\frac{\pi}{2})}. \label{eq:VT}
\end{align} 
Choosing a smaller coincidence window size $T$ generally increases the HOM visibility, but at the cost of a reduced coincidence rate. In the following, the limit $T\to\infty$ is considered, which corresponds to accepting every event. 

The correlation functions are again calculated by relating the photonic annihilation operator $a$ to the atomic collapse operator $C_2$ that corresponds to the creation of a Raman photon via the $\ket{e}\to\ket{f}$ transition, and applying the version of the quantum regression theorem \cite{Fischer_2016} for time-dependent Hamiltonians that was already mentioned in \autoref{sec:quantumjumps}. 

\section{Simulation results}\label{sec:simresults}

To investigate the statistics of back-decays during nanosecond excitation pulses and their relation with single-photon interference properties, a series of numerical calculations is performed, extracting the photon-statistical quantities $P_N$, $\langle N\rangle$, and $P_{\mathrm{Raman}}$ introduced in section \ref{sec:quantumjumps}, as well as the HOM visibility $V_{\mathrm{HOM}}=\lim_{T\to\infty}V(T)$ introduced in section \ref{sec:hom}, for various excitation parameters. 
For this, Gaussian excitation pulses with different pulse lengths $T_{\mathrm{pulse}}$ in the nanosecond range and different pulse areas $\mathcal{A}=\int\mathrm{d}t~\Omega(t)$, where $\Omega(t)$ is the time-dependent Rabi frequency, are considered. 
$T_{\mathrm{pulse}}$ is taken as the full width at half maximum (FWHM) 
of the intensity profile $\mathcal{I}(t)\propto(\Omega(t))^2$ of the 
excitation pulse. The relation $\mathcal{A} = \sqrt{\pi/(2\log2)}\,\Omega_0 T_{\mathrm{pulse}}$ holds between the pulse parameters. A detuning of $\Delta=-2\pi\times 5$\,MHz is additionally applied to closely mimic the experimental situation, in which excitation too close to resonance leads to unwanted heating of the ion. 

\begin{figure*}[t]
    \centering
    \begin{overpic}[width=0.32\linewidth]{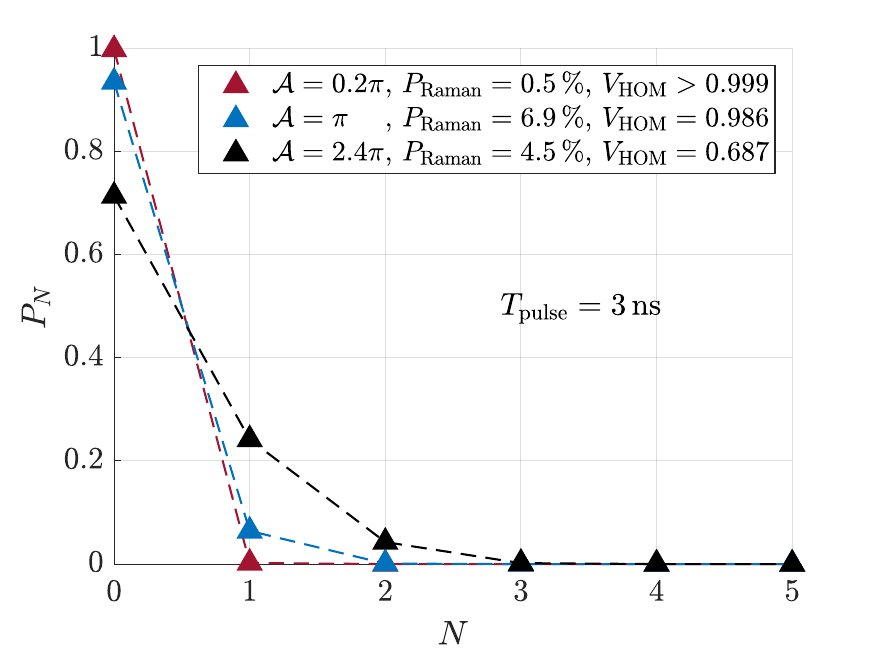}
	       	\put (2,75) {(a)}
        \end{overpic}
    \hfill
        \begin{overpic}[width=0.32\linewidth]{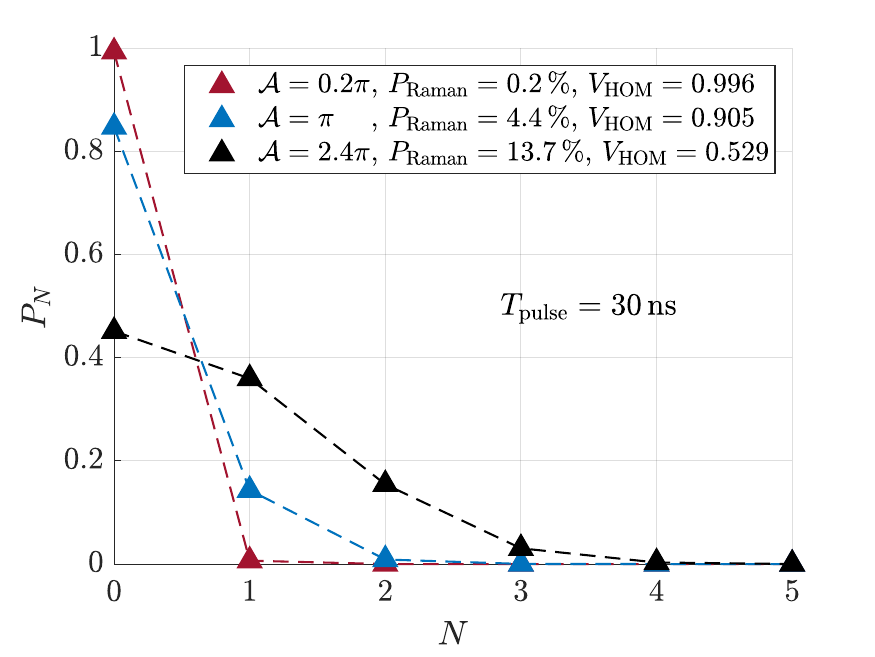}
       	\put (2,75) {(b)}
        \end{overpic}\hfill
        \begin{overpic}[width=0.32\linewidth]{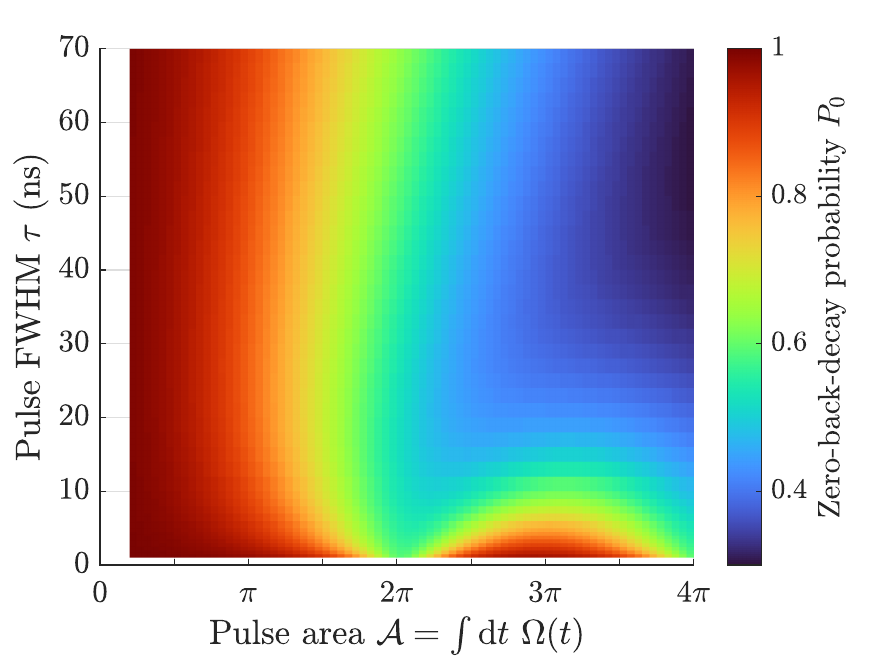}
       	\put (2,75) {(c)}
        \end{overpic}\\
        \begin{overpic}[width=0.32\linewidth]{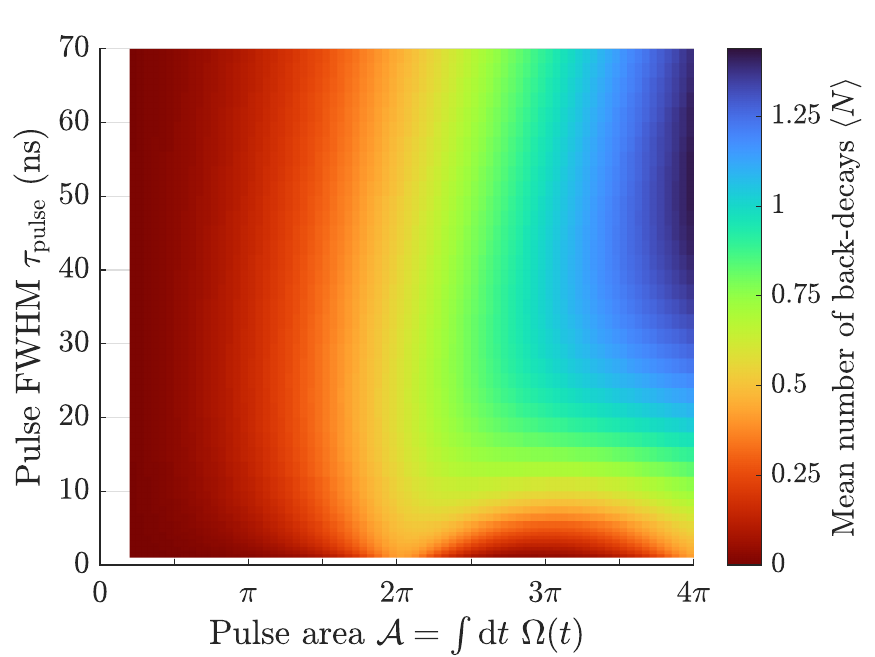}
	       	\put (2,75) {(d)}
        \end{overpic}
    \hfill
        \begin{overpic}[width=0.32\linewidth]{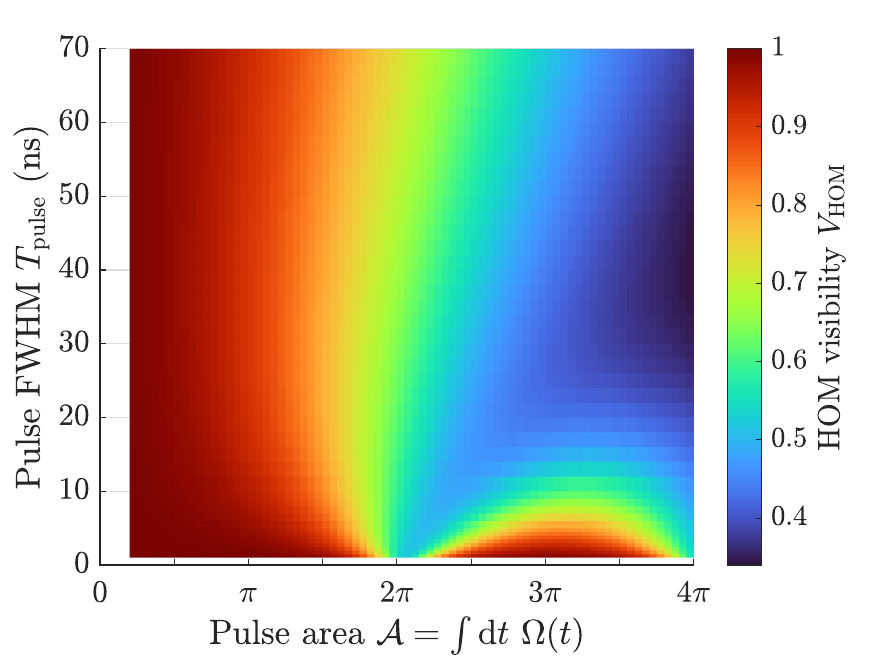}
	       	\put (2,75) {(e)}
        \end{overpic}\hfill
        \begin{overpic}[width=0.32\linewidth]{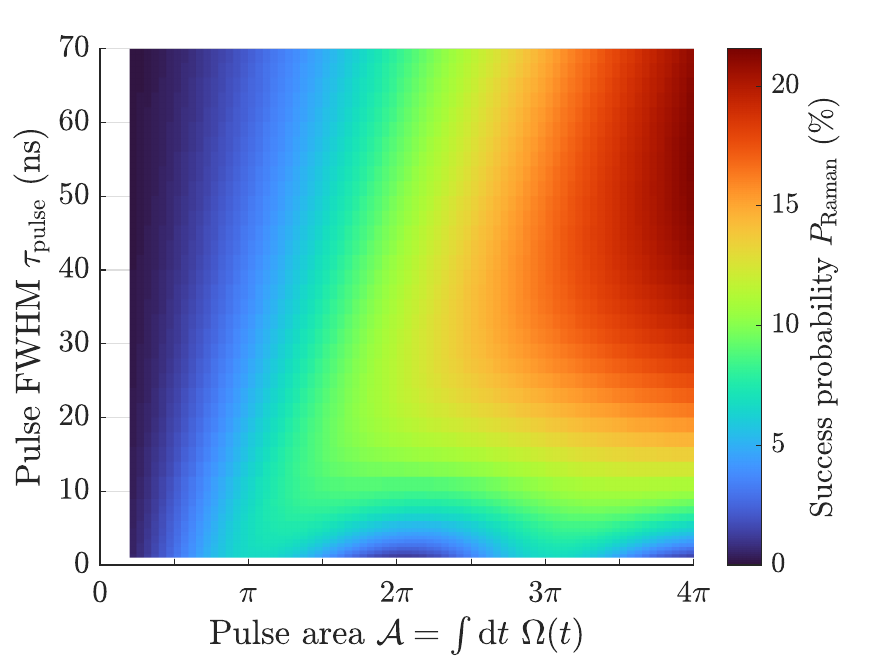}
	       	\put (2,75) {(f)}
        \end{overpic}
    \caption{Simulation results for the model outlined in \autoref{sec:model} applied to single trapped $^{40}$Ca$^+$ ions with spontaneous decay constants $\Gamma_1=2\pi\times21.49$\,MHz and $\Gamma_2=2\pi\times1.35$\,MHz $+$ $2\pi\times0.152$\,MHz. (a)--(b) Distributions of probabilities $P_N$ for exactly $N$ back-decays before the emission of a Raman photon 
    for two different pulse lengths (a) $T_{\mathrm{pulse}}=3$\,ns and (b) $T_{\mathrm{pulse}}=30$\,ns and three different pulse areas $\mathcal{A}=0.2\pi$ (red), $\pi$ (blue), and $2.4\pi$ (black) each. The data is shown as triangles with dotted lines connecting the points for visual guidance. The values of the success probability $P_{\mathrm{Raman}}$ and the HOM visibility $V_{\mathrm{HOM}}$ corresponding to each excitation pulse are listed in the legends. (c) Dependency of the zero-back-decay probability $P_0$ on the FWHM $T_{\mathrm{pulse}}$ and the pulse area $\mathcal{A}=\int\mathrm{d}t~\Omega(t)$ of the Gaussian excitation pulse used to generate a Raman photon. (d)--(f) Dependencies of (d) the mean number of back-decays $\langle N\rangle$, (e) the HOM visibility $V_{\mathrm{HOM}}$ and (f) the success probability $P_{\mathrm{Raman}}$ on the FWHM $T_{\mathrm{pulse}}$ and the pulse area $\mathcal{A}=\int\mathrm{d}t~\Omega(t)$ of the Gaussian excitation pulse used to generate a Raman photon. The data is shown as heat maps, where the mapping from numerical values to colors has been reversed for (d) to ensure that in (c)--(f), the same color always corresponds to the most desirable values.}
    \label{fig:colormaps}
\end{figure*}

To illustrate the statistics behind the number of back-decays in different excitation parameter regimes, \autoref{fig:colormaps} (a) and (b) show the distributions of the probabilities $P_N$ for two different pulse lengths $T_{\mathrm{pulse}}=3$\,ns and $T_{\mathrm{pulse}}=30$\,ns, and for different pulse areas $\mathcal{A}=0.2\pi$, $\pi$, and $2.4\pi$. These plots demonstrate two things. Firstly, the zero-back-decay probability $P_0$ is always the largest of all $P_N$, which 
corresponds to the fact that in a spontaneous emission process, early photon emission after excitation is generally more likely than later emission.
Secondly, the probabilities $P_N$ decay quickly for short or weak pulses, as multi-photon processes become improbable, while for stronger and longer pulses, this decay becomes more gradual. 
In the limit of extremely long excitation pulses, it was 
shown in \cite{Mueller_2017} that the ratio between successive values of $P_N$ is given by the branching fraction $\Gamma_1/\Gamma$, which for the photon generation scheme considered here is $\Gamma_1/\Gamma\approx 0.935$, i.e., close to unity. 
In this limit, the mean number of back-decays is $\langle N\rangle=\Gamma_1/\Gamma_2\approx 14.3$. 
Considering the standard deviation of $\langle N\rangle$, $\Delta N=(\Gamma_1/\Gamma_2)\sqrt{1+\Gamma_2/\Gamma_1}\approx 14.8$, 
one sees that while the use of long pulses has the advantage of generating the Raman photon with near-unity probability, two photons created under identical conditions will often correspond to vastly different values of $N$ and thus not overlap in time, i.e., they will not interfere. 

It is important to note that the $P_N$ are conditional probabilities and do not contain information about the likelihood of the event---the emission of a Raman photon---they are conditioned on. So a short $\pi$-pulse may show similar back-decay statistics to a longer but very weak pulse, but they have significantly different probabilities to generate the Raman photon. 

Next, the back-decay statistics are investigated in a broader parameter space, considering pulse lengths $T_{\mathrm{pulse}}$ between $1$\,ns and $70$\,ns and pulse areas $\mathcal{A}$ between $0.2\pi$ and $4\pi$. In \autoref{fig:colormaps} (c), the zero-back-decay probability $P_0$, which was introduced in \autoref{sec:quantumjumps} as a possible figure of merit for photon indistinguishability, is plotted as a heat map.
Similarly, \autoref{fig:colormaps} (d) shows the mean number of back-decays $\langle N\rangle$ before the emission of the Raman photon. 
The corresponding HOM visibility $V_{\mathrm{HOM}}$ and success probability $P_{\mathrm{Raman}}$ are plotted in \autoref{fig:colormaps} (e) and (f).
Note that the color charts of the heat maps are adapted such that the same color always corresponds to the desired situation of high photon indistinguishability or high success probability. 

One notices that the plots of \autoref{fig:colormaps} (c) to (e) display the same general behavior, verifying that the back-decay statistics of a single photon generation process, measured via $\langle N\rangle$ or $P_0$, contain information about the temporal indistinguishability of two such photons, measured via the HOM visibility. 
The success probability of \autoref{fig:colormaps} (f), however, appears to be anti-correlated with the HOM visibility. This emphasizes a trade-off between good interference properties of single photons and large photon 
generation probabilities.
\begin{figure}[h]
    \centering
    \includegraphics[width=\linewidth]{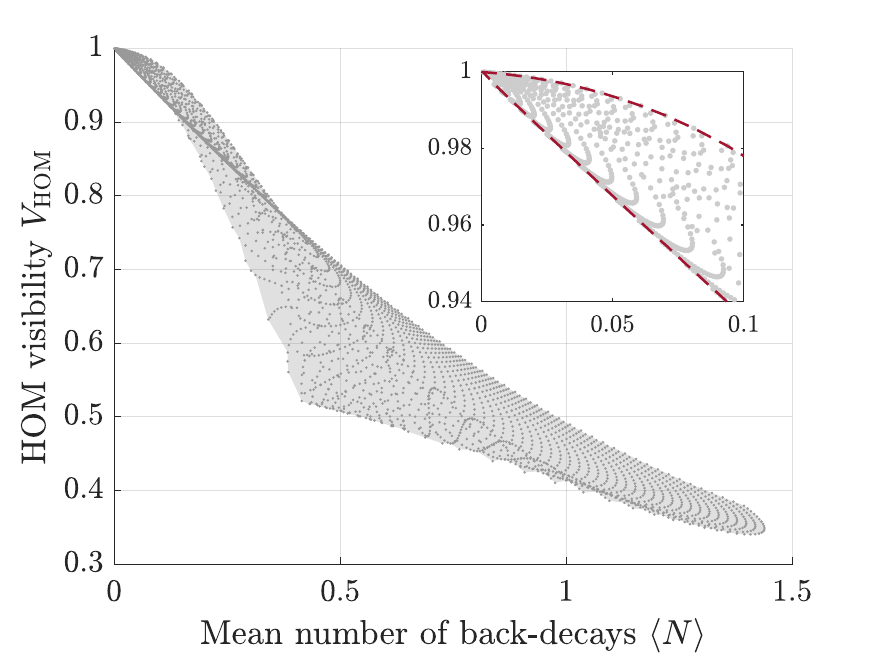}
    \caption{All pairs $(\langle N\rangle,V_{\mathrm{HOM}})$ of mean number of back-decays and HOM visibility, for all sets of excitation pulse parameters from \autoref{fig:colormaps}. The gray-shaded area approximates the region that contains all simulated data points (gray dots). Inset: Close-up of the data points for small values of $\langle N\rangle$ including two red-dotted lines showing the lower and upper bound of the HOM visibility in dependence of $\langle N\rangle$.}
    \label{fig:combined}
\end{figure}

\begin{figure}[h]
    \centering
    \includegraphics[width=\linewidth]{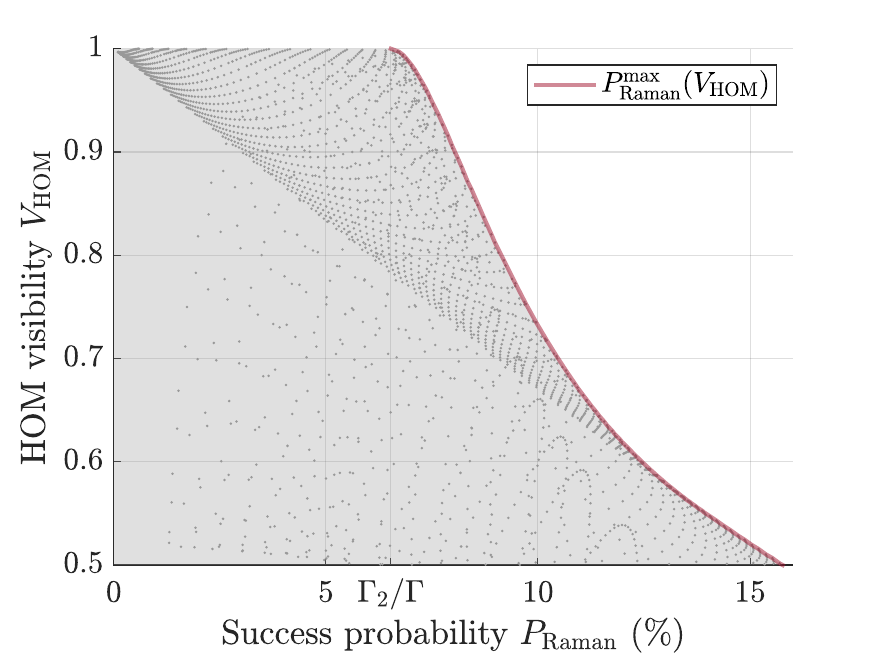}
    \caption{All pairs $(P_{\mathrm{Raman}},V_{\mathrm{HOM}})$ of success probability and HOM visibility, for all sets of excitation pulse parameters from figure \ref{fig:colormaps}. The gray-shaded area approximates the region that contains all simulated data points (gray dots). The red line indicates the boundary curve at which the success probability is maximized for a given HOM visibility.}
    \label{fig:combined2}
\end{figure}

The ability to extract information about photon interference properties from the back-decay statistics is further illustrated 
in \autoref{fig:combined}, where 
the mean number of back-decays $\langle N\rangle$ and the HOM visibility $V_{\mathrm{HOM}}$ are plotted against each other for all sets of excitation pulse parameters of \autoref{fig:colormaps}. 
While there is not a one-to-one mapping between both quantities, i.e., a given value of $\langle N\rangle$ corresponds to a range of values of  $V_{\mathrm{HOM}}$,
one observes two distinct features. First, the maximum achievable value of $V_{\mathrm{HOM}}$ decreases when $\langle N \rangle$ is increased (and vice versa). 
More importantly, in the range of high HOM visibility, its upper and lower bounds follow simple power laws. 
Numerically it is found that the bounds are well approximated by $1-0.643 \cdot \langle N\rangle \leq V_{\mathrm{HOM}} \leq 1-1.23 \cdot \langle N\rangle^{1.75}$ for $\langle N\rangle \leq 0.1$, corresponding to $0.94 \leq V_{\mathrm{HOM}} \leq 1$. 

The relationship between the success probability $P_{\mathrm{Raman}}$ and the HOM visibility $V_{\mathrm{HOM}}$ is investigated in \autoref{fig:combined2} by plotting them against each other for all excitation parameters of \autoref{fig:colormaps}. 
In this case, the pairs spread over a large area: for any realizable value of $V_{\mathrm{HOM}}$, the success probability varies between zero and a maximum, which is indicated by the red line in \autoref{fig:combined2}. This boundary marks the trade-off between the two quantities. Generally, increasing the pulse strength while decreasing the pulse length moves the photon yield closer to the maximum.
In the limit of extremely short pulses and HOM visibility close to unity, $P_{\mathrm{Raman}}$ is limited by the branching fraction $\Gamma_2/\Gamma\approx 6.5$\,\%, corresponding to the photon generation probability with an instantaneous $\pi$-pulse. 
If the pulses are longer, the possibility of back-decay and subsequent re-excitation allows one to push the success probability beyond the limit set by the branching fraction, but at the cost of reduced HOM visibility. 

\begin{figure}[h]
    \centering
    \includegraphics[width=\linewidth]{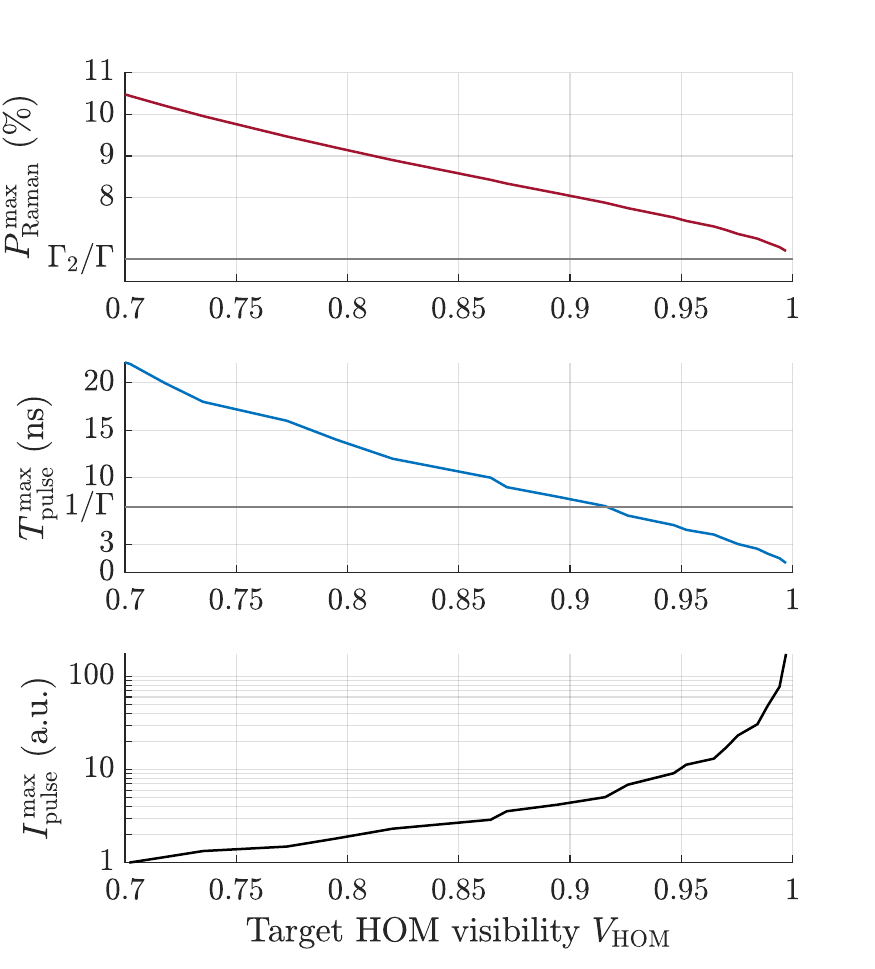}
    \caption{The three graphs show, for a given target HOM visibility, the
    maximum achievable success probability $P_{\mathrm{Raman}}^{\,\mathrm{max}}$ (top), as well as the pulse length $T_{\mathrm{pulse}}^{\,\mathrm{max}}$ (middle) and pulse intensity $I_{\mathrm{pulse}}^{\,\mathrm{max}}$ (bottom) that realize this value.}
    \label{fig:boundary}
\end{figure}

The maximum photon generation probability for a given target HOM visibility, $P_{\mathrm{Raman}}^{\mathrm{max}}$, marked by the red curve in \autoref{fig:combined2}, is reached for certain optimum excitation parameters, a pulse length $T_{\mathrm{pulse}}^{\,\mathrm{max}}$ and a pulse area $\mathcal{A}^{\,\mathrm{max}}$. 
In \autoref{fig:boundary}, the maximum success probability, as well as the corresponding pulse length and pulse intensity $I_{\mathrm{pulse}}^{\,\mathrm{max}}\propto (\mathcal{A}^{\,\mathrm{max}}/T_{\mathrm{pulse}}^{\,\mathrm{max}})^2$, are plotted against the HOM visibility. 
With a pulse width comparable to the excited state lifetime of $\sim 7$\,ns, a value $V_\mathrm{HOM} \gtrsim 0.92$ may be reached at moderate pulse intensity. Up to $V_\mathrm{HOM} \approx 0.99$, the required pulse width decreases linearly to $\sim 2$\,ns, and the intensity rises by one order of magnitude. 
Only as the desired HOM visibility approaches unity, the excitation requires sub-nanosecond pulse lengths and exceedingly large power, which is a main disadvantage of working in this regime. 

\section{Two emitters}

In the previous section, it was established that there is a trade-off between photon generation probability and photon indistinguishability, and that there exists an optimal photon yield for a given target HOM visibility. So far, however, excitation schemes beyond a single pulse have not been considered, and the possibility of improving the HOM visibility by limiting the coincidence window size has not been discussed either. These additional measures are considered in the following and applied to the specific scenario of performing dual-rail photonic entanglement swapping between two emitters. There is a direct connection to the considerations of the previous sections in that the indistinguishability of single photons, expressed through $V_{\mathrm{HOM}}$, limits the achievable fidelity $F$ of an entanglement swapping operation via the relation $F=(1+V_{\mathrm{HOM}})/2$ \cite{Craddock_2019}. We therefore keep the HOM visibility as a figure of merit also in this modified scenario.

The other figure of merit, rather than the success probability to generate a single photon, is the joint probability of finding two photons with orthogonal polarizations at the detectors behind a beam splitter, since the detection of such a coincidence event constitutes the Bell state measurement that establishes entanglement between the emitters \cite{Simon_2003}. This coincidence probability is of the form
\begin{align}
    P^{(1)}_{\mathrm{coinc}}(T) = (P_{854})^2\cdot \mathcal{W}(T), 
\end{align}
where 
\begin{align}
    \mathcal{W}(T)=\frac{\int_0^\infty \mathrm{d}t~\int_{-T/2}^{T/2}\mathrm{d}\tau~G_{\mathrm{HOM}}^{(2)}(t,\tau,\frac{\pi}{2})}{\left(\int_0^\infty\mathrm{d}t~\langle a_2^\dagger(t)a_2(t)\rangle \right)^2}\leq \frac{1}{2}
\end{align}
is a window function that gives the fraction of coincidence events depending on the coincidence window size $T$. Decreasing the coincidence window size is one method (below also termed time-filtering) to increase HOM visibility without changing the excitation pulse parameters, but it comes at the cost of a reduced coincidence probability \cite{Walker_2020, Meraner_2020}.

Before the effect of time-filtering is discussed, a method to increase the coincidence probability without altering the HOM visibility is introduced, namely the use of a train of successive excitation pulses before the ions are reset to the ground state. In the limit of infinitely many pulses, this will completely empty the ground state populations, resulting in unit probability to generate a Raman photon from both emitters. Due to the parasitic decay to $\mathrm{D}_{3/2}$, the probability of generating a photon at 854\,nm is then $P_{854} = \Gamma_{854}/(\Gamma_{854}+\Gamma_{850}) \approx 90$\,\%.
Coincidence, however, requires the two photons to be generated by the same excitation pulse in the train. The coincidence probability after $n_{\mathrm{pulses}}$ excitation pulses 
is therefore given by 
\begin{align}
    P^{(n_{\mathrm{pulses}})}_{\mathrm{coinc}}(T) = P^{(1)}_{\mathrm{coinc}}(T)\cdot\sum_{n=1}^{n_{\mathrm{pulses}}}(1-P_{\mathrm{Raman}})^{2(n-1)},
\end{align}
which for $n_{\mathrm{pulses}}\to\infty$ becomes
\begin{align}    
    P_{\mathrm{coinc}}^{(\infty)}(T) := \frac{P^{(1)}_{\mathrm{coinc}}(T)}{1-(1-P_{\mathrm{Raman}})^2}.
\end{align}
Here we assume that the excitation pulses are well separated in time, such that for both emitters the population of the ground state is reduced by a factor $(1-P_{\mathrm{Raman}})$ with each successive excitation pulse. 

When applying a series of pulses to enhance the photon yield, the central result from \autoref{sec:simresults} regarding the trade-off between HOM visibility and success probability---now considering the coincidence probability as the figure of merit---remains valid: the same excitation pulse parameters result in the maximal photon yield for a given target HOM visibility. 

\begin{figure}[h]
    \centering
    \begin{overpic}[width=\linewidth]{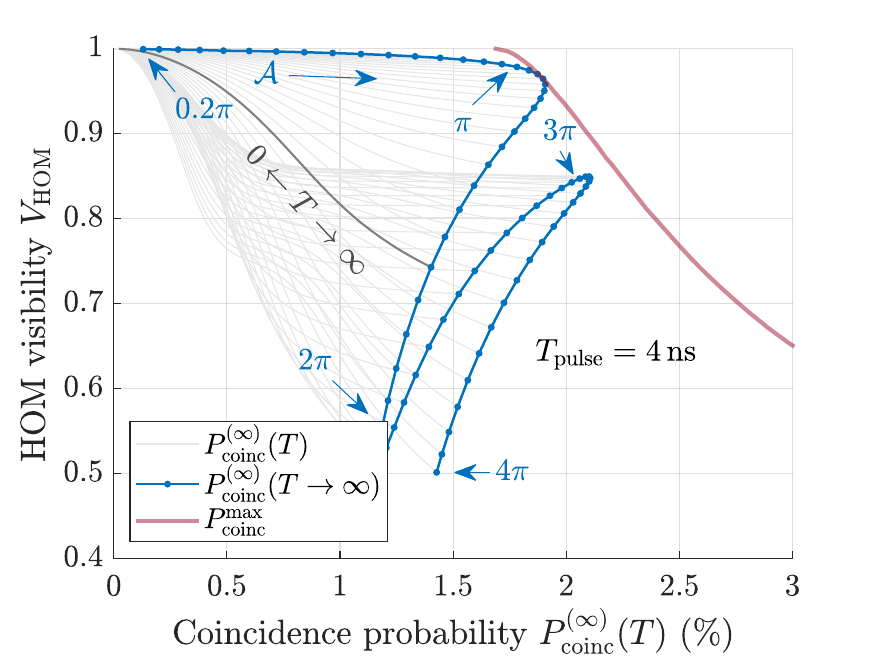}
	       	\put (2,70) {(a)}
        \end{overpic} \\ 
        \begin{overpic}[width=\linewidth]{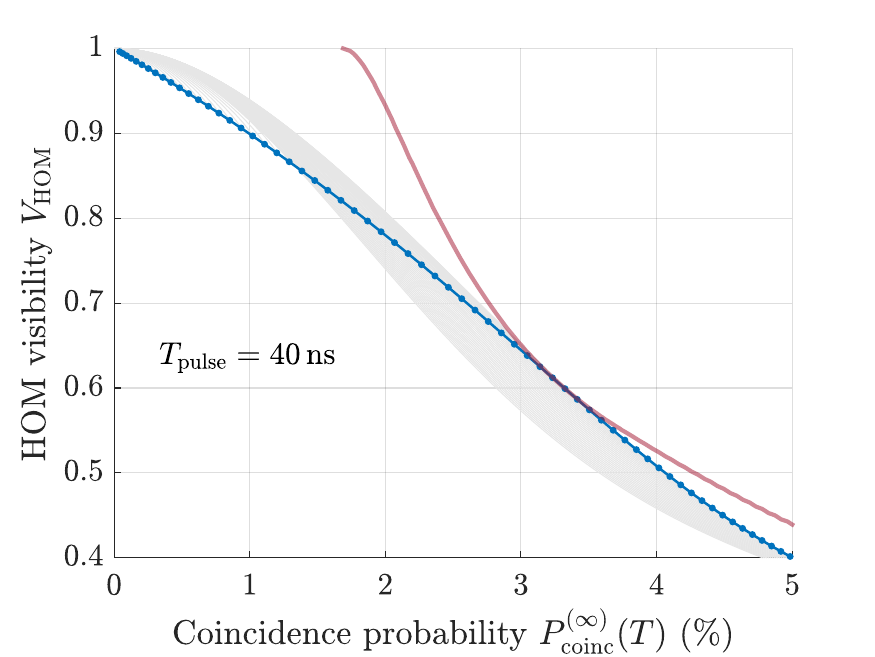}
	       	\put (2,70) {(b)}
        \end{overpic}
    \caption{Effect of time-filtering. Value pairs $(P_{\mathrm{coinc}}^{(\infty)}, V_{\mathrm{HOM}})$ realizable by varying the pulse area $\mathcal{A}$ and coincidence window size $T$ for a fixed pulse length of (a) $T_{\mathrm{pulse}}=4$\,ns and (b) $T_{\mathrm{pulse}}=40$\,ns. 
    Blue points (connected by a blue line for visual guidance) mark pairs   $(P_{\mathrm{coinc}}^{(\infty)}(T),V_{\mathrm{HOM}}(T))$ for $T\to \infty$ and $\mathcal{A}$ varying between $0.2\,\pi$ and $4\,\pi$ in steps of $0.05\,\pi$. Gray curves are pairs $(P_{\mathrm{coinc}}^{(\infty)}(T),V_{\mathrm{HOM}}(T))$ for fixed $\mathcal{A}$, and $T$ varying from 0 to $\infty$. 
    In (a), the curve for $\mathcal{A}=1.75\,\pi$ is plotted in darker gray to illustrate how it evolves from $T=0$ to $T \to \infty$. In both (a) and (b), the red line shows the maximum achievable coincidence probability $P_{\mathrm{coinc}}^{\mathrm{max}}$ for any given HOM visibility.}
    \label{fig:timefiltering}
\end{figure}

In the following, we assume an infinite train of excitation pulses and investigate how time-filtering can be used to optimize the photon yield for a given target HOM visibility, finding that in some cases it enables a higher coincidence probability than what can be achieved by simply varying the pulse strength.
This is demonstrated in \autoref{fig:timefiltering} for two example pulse length values, one shorter than the excited state lifetime (4\,ns) and one longer (40\,ns). Plotted are value pairs $(P_{\mathrm{coinc}}^{(\infty)}(T),V_{\mathrm{HOM}}(T))$ that result by varying the single-pulse area $\mathcal{A}$ and coincidence window size $T$, for an infinite train of pulses. 
Fixing also $\mathcal{A}$ and varying only $T$ results in the gray curves $T\mapsto (P_{\mathrm{coinc}}^{(\infty)}(T),V_{\mathrm{HOM}}(T))$ that start at the point $(0,1)$ for $T=0$. At their endpoints, $T\to\infty$, the full temporal extent of the photon resulting from each excitation pulse is taken into account.  
The endpoints are drawn as blue dots; the blue line serves for visual guidance, but also shows how $(P_{\mathrm{coinc}}^{(\infty)}(T),V_{\mathrm{HOM}}(T))$, for $T\to\infty$, varies with the pulse area in the considered range. 

For $T_{\mathrm{pulse}}=4$\,ns, given some fixed value of the HOM visibility, the point with the highest coincidence probability always lies on the blue line, i.e. it is realizable without the use of time-filtering. In contrast, for $T_{\mathrm{pulse}}=40$\,ns, if $V_{\mathrm{HOM}}\gtrsim0.65$ is targeted, the photon yield may be optimized by using time-filtering, which is apparent from the fact that parts of the curves $T\mapsto (P_{\mathrm{coinc}}^{(\infty)}(T),V_{\mathrm{HOM}}(T))$ (gray lines) lie to the right of the end points. 
One final thing to note is that none of the points including time-filtering lie to the right of the boundary curve  according to \autoref{fig:combined2}, marked by the red line. 
This means that unless one is restricted to pulse lengths beyond the excited-state lifetime, time-filtering is not necessary to achieve the highest photon yield for a given target HOM visibility; rather there exists some excitation pulse that will maximize the photon yield without the use of time-filtering.

\section{Other excitation schemes and ion species}\label{sec:species}

The $\mathrm{S}_{1/2}\to\mathrm{P}_{3/2}\to\mathrm{D}_{5/2}$ photon generation scheme is one of several ways to generate single photons from a trapped $^{40}$Ca$^+$ ion, and is favored in practice because the 854-nm emission wavelength converts easily to the telecom band. 
The inverted scheme, $\mathrm{D}_{5/2}\to\mathrm{P}_{3/2}\to\mathrm{S}_{1/2}$, is advantageous in terms of success probability and photon indistinguishability, but converting the near-UV emission at 393\,nm to the telecom band for long-distance applications is challenging and achieves lower efficiencies than for 854\,nm \cite{Liu_2026}. Additionally, the single-photon wavelength of 854\,nm allows for integration of an optical cavity to enhance the emission \cite{Stute_2012}, while in contrast, for cavities near the UV, rapid degradation has been observed under experimental conditions \cite{Ballance_2017}.

An alternative approach prepares the ion in $\mathrm{D}_{3/2}$ and generates 854\,nm photons via $\mathrm{D}_{3/2}\to\mathrm{P}_{3/2}\to\mathrm{D}_{5/2}$. This scheme has been proposed as a route to high HOM visibility \cite{Bergerhoff_2021_MT, Cai_2025_QOXI}, and the 850-nm excitation wavelength is more convenient than near-UV light. The large branching fraction into $\mathrm{S}_{1/2}$, however, substantially reduces the probability of generating an 854-nm photon, and the required additional state preparation steps lower the achievable repetition rate. 

The above considerations and trade-offs apply similarly to schemes using $\mathrm{P}_{1/2}$ as the excited state, such as $\mathrm{S}_{1/2} \to \mathrm{P}_{1/2} \to \mathrm{D}_{3/2}$ which has been realized with  $^{40}$Ca$^+$ in \cite{Cui_2025}.

Two other alkaline-earth trapped-ion species, $^{88}$Sr$^+$ and $^{138}$Ba$^+$, have been established as platforms for quantum networking and computing \cite{Wright_2018, Hannegan_2022, Crocker_2019}. Here, all schemes realized or proposed so far go through $\mathrm{P}_{1/2}$ as the excited state. The $\mathrm{D}_{3/2} \to \mathrm{P}_{1/2} \to \mathrm{S}_{1/2}$ scheme has been implemented with $^{138}$Ba$^+$ \cite{Hannegan_2022, Crocker_2019} and would be suitable for $^{88}$Sr$^+$ with frequency conversion from 422\,nm to the telecom C-band \cite{Wright_2018}, while its inverse has been realized with $^{88}$Sr$^+$ \cite{Zalewski_2026}. 
Finally, a scheme involving $\mathrm{D}_{3/2}$, $\mathrm{P}_{3/2}$, and $\mathrm{D}_{5/2}$ in $^{171}$Yb$^+$ with direct emission at 1650\,nm has also been proposed \cite{Wang_2020}.

It is straightforward to apply the numerical calculations discussed in \autoref{sec:simresults} to the other ion species and photon generation schemes by altering the initial states, collapse operators, and spontaneous decay rates (listed for all relevant ion species and energy levels in appendix \ref{appendix:rates}). In order to assess the performance of an ion species with respect to remote entanglement generation based on photonic Bell state measurements, it is important to additionally consider wavelength-dependent losses in optical fibers. For implementations over large distances, the emission wavelength should ideally be close to the telecom band or be frequency-converted to it. 
\autoref{tab:species} lists photon generation schemes based 
on $\Lambda$-type Raman transitions in trapped ions, that have been either implemented or proposed. The table shows the respective emission wavelengths, the target wavelengths after conversion (if implemented; otherwise, the emission wavelength is considered the target wavelength), the conversion efficiencies, and the losses at the target wavelengths in SMF-28e+ fiber \cite{SMF28}. 

\begin{figure}[h]
    \centering
    \includegraphics[width=\linewidth]{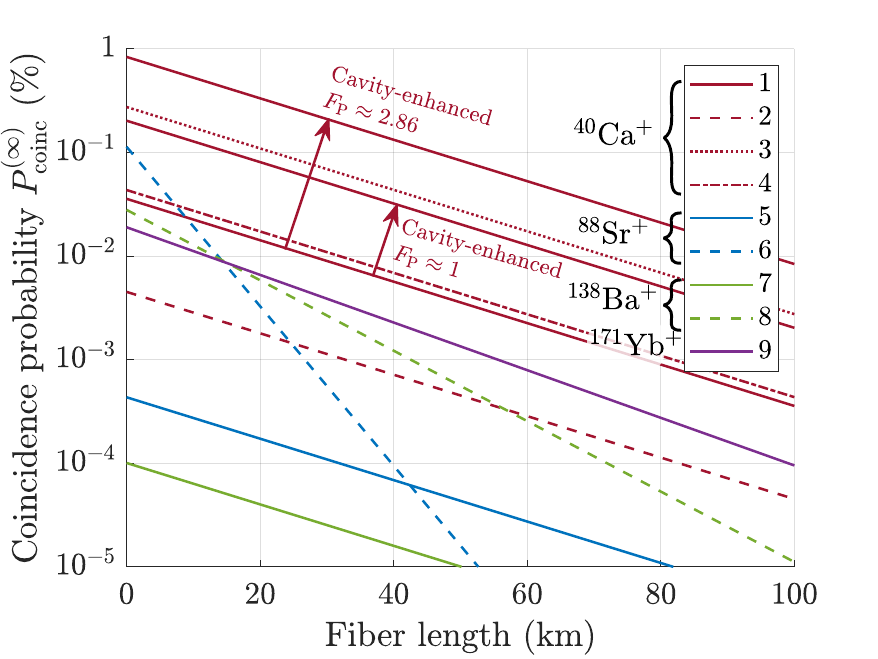}
    \caption{Coincidence probability in dependence of SMF-28e+ fiber length for different ion species and single Raman photon generation schemes based on reported conversion efficiencies and fiber losses according to \cite{SMF28,Zalewski_2026}. The photon generation schemes corresponding to 
    the numbers 1--9 in the legend are listed in \autoref{tab:species}. Two arrows connect to the expected cavity-enhanced coincidence probability for the first scheme, with two assumed values of the Purcell enhancement, $F_{\mathrm{P}}\approx 1$ and $F_{\mathrm{P}}\approx 2.86$.}
    \label{fig:species}
\end{figure}

Based on these specifications, the coincidence probability assuming excitation with a train of infinitely many instantaneous $\pi$-pulses is plotted as a function of total fiber length (i.e., the sum of the lengths both photons travel) in \autoref{fig:species}. For these plots it was assumed that a fraction of 20\,\% of the spontaneously emitted photons is collected. This value is based on the setup reported in \cite{Carter_2024}, which to our knowledge achieves the highest free-space photon collection efficiency to date for trapped-ion quantum networking, using collection optics with a numerical aperture (NA) of 0.8. The results aim to compare the expected performance of state-of-the-art ion trap systems in a dual-rail entanglement swapping application over large distances under identical boundary conditions. It should be noted that experimental demonstrations of this application have not necessarily been reported for all schemes or species considered here.

\begin{table*}\centering
\ra{1.3}
\begin{tabular}{@{}cclcccl@{}}\toprule
~~\#~~ & 
\multicolumn{2}{c}{Excitation scheme} & Wavelength (nm) & QFC eff. (\%) & losses (dB/km) & References\\ 
\midrule
$1$ & 
$^{40}$Ca$^+$ & $\mathrm{S}_{1/2}\to\mathrm{P}_{3/2}\to\mathrm{D}_{5/2}$ & $854\to 1550$ & 57.2 & 0.2 & \cite{Bock_2024,Bergerhoff_2026}\\
$2$ & 
& $\mathrm{D}_{3/2}\to\mathrm{P}_{3/2}\to\mathrm{D}_{5/2}$ & $854\to 1550$ & 57.2 & 0.2 & \cite{Bergerhoff_2021_MT, Cai_2025_QOXI}\\
3 & & $\mathrm{D}_{5/2}\to\mathrm{P}_{3/2}\to\mathrm{S}_{1/2}$ & $393\to 1550$ & 28  & 0.2 & \cite{Liu_2026}\\
4 & & $\mathrm{S}_{1/2}\to\mathrm{P}_{1/2}\to\mathrm{D}_{3/2}$  & $866\to 1550$ & 57.2 & 0.2 & \cite{Cui_2025}\\
5 & $^{88}$Sr$^+$ & $\mathrm{D}_{3/2}\to\mathrm{P}_{1/2}\to\mathrm{S}_{1/2}$ & $422\to 1561$ & 1.1  & 0.2 & \cite{Wright_2018}\\
6 & & $\mathrm{S}_{1/2}\to\mathrm{P}_{1/2}\to\mathrm{D}_{3/2}$  & 1092 & - & 0.77 & \cite{Zalewski_2026}\\
7 & $^{138}$Ba$^+$ & $\mathrm{D}_{3/2}\to\mathrm{P}_{1/2}\to\mathrm{S}_{1/2}$  & $493\to 1534$ & 0.66 & 0.2 & \cite{Hannegan_2022, Crocker_2019,Hannegan_2021}\\
8 & & $\mathrm{D}_{3/2}\to\mathrm{P}_{1/2}\to\mathrm{S}_{1/2}$ & $493\to 1287$ & 11  & 0.34 & \cite{Saha_2023}\\
9 & $^{171}$Yb$^+$ &$\mathrm{D}_{3/2}\to\mathrm{P}_{3/2}\to\mathrm{D}_{5/2}$ & 1650 & -& 0.25 & \cite{Wang_2020}\\
\bottomrule
\end{tabular}
\caption{Implemented or proposed single-photon generation schemes with initial and target wavelengths, conversion efficiencies (if implemented), and losses in SMF-28e+ fiber at the target wavelengths. The conversion efficiency from 866\,nm to 1550\,nm has not been directly reported but is assumed to be equal to the efficiency from 854\,nm to the same target wavelength due to the proximity of the two initial wavelengths. The numbering in the first column is the same as in \autoref{fig:species}. 
In the references, from which the reported conversion efficiencies are taken, the respective single-photon schemes are implemented or proposed, but not necessarily for dual-rail entanglement swapping.}
\label{tab:species}
\end{table*}

It is apparent that the schemes which would perform best 
in this comparison are those based on $^{40}$Ca$^+$. Only for less than 10\,km of fiber, a $^{88}$Sr$^+$-based scheme appears among the top three. This is, however, mainly a consequence of the absence of conversion; for these distances, a fair comparison would also incorporate other schemes without conversion. 
Among the three Calcium-based schemes which would perform best over large distances, an implementation with the specifications from \cite{Liu_2026} would outperform the others by almost an order of magnitude. This is owed to the more favorable branching fraction to the final state outweighing the lower conversion efficiency when compared to the specifications of the inverted scheme \cite{Bock_2024}.

An obvious measure to enhance the single-photon yield would be to modify the decay rate on the relevant transition by the Purcell effect in an optical cavity. As an example, we consider the $\mathrm{S}_{1/2}\to\mathrm{P}_{3/2}\to\mathrm{D}_{5/2}$ scheme used in our implementations and assume a Purcell enhancement of $F_{\mathrm{P}}\approx 1$ for the 854-nm emission. The calculated coincidence probability for this setup is also included in \autoref{fig:species}. It is expected to lead to a similar performance as the best free-space scheme. To determine the coincidence probability for the cavity-enhanced scheme, a modified branching fraction of the 854-nm transition according to $\Gamma_{854} \to (1+F_{\mathrm{P}}) \Gamma_{854}$, a fraction $F_{\mathrm{P}} / (1+F_{\mathrm{P}})$ of photons emitted into the cavity mode, and a 2:1 ratio of the mirror transmissions were considered. These parameters are taken from our ongoing efforts to construct a new ion trap with an integrated sub-mm fiber-based optical cavity \cite{Becker2025_conference}. Since the Purcell enhancement scales with the finesse of the cavity, even stronger enhancement is possible via optimization of its manufacturing and handling. According to the manufacturer-specified mirror transmissions of $100$\,ppm and $200$\,ppm for the setup of \cite{Becker2025_conference}, a maximal finesse of $\sim 20900$ is theoretically possible, which would increase the Purcell enhancement to $F_{\mathrm{P}}\approx 2.86$. The expected coincidence probability for this scenario is also included in \autoref{fig:species} and shows a significant enhancement compared to the best of the considered free-space realizations.

\section{Summary and Discussion}

With numerical simulations, which are backed by experimental results in \cite{Baumgart_2026exp}, we have shown that when driving a $\Lambda$-type Raman transition to generate single photons, spontaneous decays back to the initial state have a major influence on the interference capabilities of the photons. This is particularly pronounced when excitation pulses with pulse lengths on the order of the excited state lifetime are used. 
These back-decays also cause photons from two otherwise identical emitters to become temporally distinguishable, which affects the fidelity of photonic Bell measurements used, for example, in entanglement swapping. 
The statistics of back-decays, with their mean number $\langle N\rangle$ being an experimentally accessible single-emitter quantity \cite{Baumgart_2026exp}, 
allow one to estimate upper and lower bounds for the HOM interference visibility expected for photons from two identical emitters excited with the same pulse. 

We studied in detail how HOM visibility and success probability depend on the excitation scheme and parameters, and we characterized the trade-off between them. Generally, HOM visibility may be increased by sacrificing photon yield. 
If one fixes a target HOM visibility, the photon yield is limited to a maximum. If the targeted HOM visibility is close to 1, this maximum is set by the branching fraction of the excited state. Reaching it, however, requires pulse lengths much shorter than the excited state lifetime and a disproportionately higher power compared to excitation pulses in the nanosecond regime, which can still achieve high HOM visibilities at moderate powers and with a higher photon yield.

Two additional approaches that optimize single-photon generation with respect to the trade-off between photon yield and indistinguishability were discussed for application to dual-rail photonic entanglement swapping with two emitters. The use of a train of multiple excitation pulses increases the photon yield without affecting the indistinguishability or entanglement fidelity. The use of time-filtering, i.e., only accepting two-photon coincidences within a finite time-window, turns out to be beneficial, but only for excitation pulse lengths considerably larger than the excited-state lifetime. Otherwise, the photon yield for a given target HOM visibility is always maximized by optimizing the excitation pulse length and pulse area.

Finally, the performance of different trapped-ion species ($^{40}$Ca$^+$, $^{88}$Sr$^+$, $^{138}$Ba$^+$, and $^{171}$Yb$^+$) and different proposed or implemented photon generation schemes 
was compared with respect to long-distance entanglement swapping. This comparison was based on 
free-space photon collection and reported conversion efficiencies to wavelengths in (or close to) the telecom band.
The three schemes that showed the best potential over fiber distances of more than 10\,km are all based on $^{40}$Ca$^+$, one main reason for this being the availability of high-efficiency quantum frequency converters. 

A promising next step towards scaling up the performance of trapped-ion platforms in quantum networks is the integration of optical cavities. Various ion-cavity experiments were realized in the past, for example with $^{40}$Ca$^+$ on several different lines \cite{Mundt_2002, Keller2004SinglePhotons, Stute_2012}, with $^{174}$Yb$^+$ in the IR \cite{Steiner_2014} and the UV \cite{Ballance_2017}, and with $^{88}$Sr$^+$ at a blue wavelength \cite{Leibrandt_2009}.
More recent experiments have been aimed towards quantum networking, using $^{40}$Ca$^+$ in a centimeter-scale optical cavity \cite{Meraner_2020, Krutyanskiy_2023_QR}. Photon generation in these works is based on the $\mathrm{S}_{1/2} \to \mathrm{P}_{3/2} \to \mathrm{D}_{5/2}$ scheme, which is also used in our own experimental demonstrations \cite{Bock_2018, Bergerhoff_2024, Baumgart_2026exp}, and on which the discussion in the main part of this manuscript is based.
We are currently assembling an ion trap with an integrated sub-millimeter cavity \cite{Becker2025_conference}, which is expected to improve the performance of our current single-photon generation by close to an order of magnitude. Ideally, the short cavity combines high Purcell enhancement with a ring-down time below the atomic lifetime, such that both photon generation probability and generation rate are enhanced. This would open up the way for multi-ion multiplexing, building on promising recent experimental demonstrations \cite{Canteri_2025, You_2026, Cui_2025}. 

\section*{Author contributions}
P.\,B. performed the simulations. P.\,B. wrote the manuscript with input from all authors. M.\,B. contributed to the research and writing of the section ''Other excitation schemes and ion species'' and provided scientific input throughout the project. J.\,E. conceived, coordinated, and supervised the research.

\begin{acknowledgments}
We acknowledge support from the Federal Ministry of Research, Technology and Space (BMFTR) through projects Q.sync (16KISQ045), QR.X (16KISQ001K), and QR.N (16KIS2180). 
\end{acknowledgments}

\appendix
\section*{Appendix}

\section{Details of the HOM-visibility calculation}\label{appendix:details}
For the calculation of the HOM visibility, one considers the two incident fields 
$E_{a,b}(t)= E_{a,b}^{(+)}(t)+E_{a,b}^{(-)}(t)$ at the spatial input modes $a$ and $b$ of a 50:50 beam splitter, which may be written in a dimensionless form as 
\begin{align}
    E_a^{(+)}(t)&= i a(t)~\hat{e}_x,\\
    E_b^{(+)}(t)&= ib(t)(\cos\phi~\hat{e}_x + \sin\phi~\hat{e}_y), \\
    E_{a,b}^{(-)}(t)&=\left(E_{a,b}^{(+)}(t)\right)^\dagger.
\end{align}
Without loss of generality, the polarization unit vector of the field at mode $a$ is taken to be the unit vector $\hat{e}_x$ in the $x$-direction, while the polarization unit vector at mode $b$ is taken to be a superposition of $\hat{e}_x$ 
and the unit vector $\hat{e}_y$ in the $y$-direction, with $\phi$ being the angle between the two polarization vectors. 

The two output modes of the beam splitter are labeled $c$ and $d$, and the corresponding electric field operators read 
\begin{align}
    E_c^{(+)}(t)&=\frac{1}{\sqrt{2}}\left(
    E_a^{(+)}(t)+iE_b^{(+)}(t)
    \right),\label{eq:field_out1}\\
    E_d^{(+)}(t)&=\frac{1}{\sqrt{2}}\left(
    iE_a^{(+)}(t)+E_b^{(+)}(t)
    \right). \label{eq:field_out2}
\end{align}
With this, the normally ordered two-time second-order correlation function 
\begin{align}
    G^{(2)}_{\mathrm{HOM}}(t,\tau,\phi)=\langle 
    E_c^{(-)}(t)E_d^{(-)}(t+\tau)E_d^{(+)}(t+\tau)
    E_c^{(+)}(t)
    \rangle, \label{eq:G2HOM}
\end{align}
of the output fields is expressed in terms of correlation functions of the input fields as 
\begin{align}
    G^{(2)}_{\mathrm{HOM}}&(t,\tau,\phi)=\frac{1}{4}\bigl(G_a^{(2)}(t,\tau)+G_b^{(2)}(t,\tau) \nonumber\\
    &+\langle n_a(t)\rangle\langle n_b(t+\tau)\rangle  
    +\langle n_a(t+\tau)\rangle\langle n_b(t)\rangle\nonumber \\
    &-2\cos^2\phi~\mathfrak{Re}\{(G_a^{(1)}(t,\tau))^*G_b^{(1)}(t,\tau)\}\bigr), 
\end{align}
where $^*$ denotes complex conjugation and 
\begin{align}
    \langle n_a(t)\rangle &= \langle a^\dagger(t)a(t)\rangle,\\
    G_{a}^{(1)}(t,\tau) &= \langle a^\dagger(t)a(t+\tau)\rangle, \\
    G_a^{(2)}(t,\tau)&=\langle a^\dagger(t)a^\dagger(t+\tau)a(t+\tau)a(t)\rangle, 
\end{align}
and equivalently for $b$ \cite{Woolley_2013}. With this formalism, the general case of photons with different temporal profiles, resulting from different temporal envelopes $f(t)$ of the driving fields, may be calculated. If the temporal profiles are the same for both inputs, the expression simplifies to 
\begin{align}
    G^{(2)}_{\mathrm{HOM}}&(t,\tau,\phi)=\frac{1}{2}\bigl(G_a^{(2)}(t,\tau)
    +\langle n_a(t)\rangle\langle n_a(t+\tau)\rangle \nonumber\\
    &-\cos^2\phi~\vert G_a^{(1)}(t,\tau))\vert^2\bigr), 
\end{align}
since all correlation functions of $a$ and $b$ are the same for identical emitters. The expression is simplified even further if one takes into account that at most one Raman photon is scattered from the $\Lambda$-system before resetting to the ground state, resulting in $G^{(2)}_a(t,\tau)=0$, which leads to the final expression in \autoref{eq:G2hom}.

\section{Validity of the three-level model}\label{appendix:threelevel}

The modeling of $^{40}$Ca$^+$ (and other trapped-ion species) as an effective three-level system is valid for most practical implementations, where the presence of multiple excitation and decay channels may however lead to the necessity of filtering out certain single-photon detection events. To give an example, in the experiments carried out in \cite{Baumgart_2026exp}, the ions are initially always in a mixture of both magnetic sub-states of $\mathrm{S}_{1/2}$, such that both the $\ket{\mathrm{S}_{1/2},-\frac{1}{2}}\to\ket{\mathrm{P}_{3/2},-\frac{1}{2}}$ and the $\ket{\mathrm{S}_{1/2},+\frac{1}{2}}\to\ket{\mathrm{P}_{3/2},+\frac{1}{2}}$ transitions are simultaneously excited by the linearly polarized 393-nm laser. From each of the sub-states of the excited state, three decay paths to $\mathrm{D}_{5/2}$ are possible, one of which emits a $\pi$-polarized 854-nm photon that is however not collected since the collection optics are aligned with the quantization axis. 
If it is necessary to only consider coincidence events that resulted from spontaneous decay from the same sub-state of $\mathrm{P}_{3/2}$, as is the case for entanglement swapping \cite{Bergerhoff_2026}, this can be attained via state discrimination in the atomic readout.
Such post-selection reduces the number of usable coincidence events but also ensures the validity of the three-level approximation by singling out the respective excitation and decay channels. In \cite{Bergerhoff_2026}, those were $\ket{\mathrm{S}_{1/2},-\frac{1}{2}}\to\ket{\mathrm{P}_{3/2},-\frac{1}{2}}\to\ket{\mathrm{D}_{5/2},-\frac{3}{2}}$ for right-circularly polarized photons and $\ket{\mathrm{S}_{1/2},-\frac{1}{2}}\to\ket{\mathrm{P}_{3/2},-\frac{1}{2}}\to\ket{\mathrm{D}_{5/2},+\frac{1}{2}}$ for left-circularly polarized photons, which are treated independently and irrespective of their frequency difference by the model. 

Another consequence of using a three-level model is that the metastable levels $\mathrm{D}_{3/2}$ and $\mathrm{D}_{5/2}$ are combined into a singular final level $\ket{f}$. This is done by adding up the decay constants for the two transitions $\mathrm{P}_{3/2} \to \mathrm{D}_{5/2}$ and $\mathrm{P}_{3/2} \to \mathrm{D}_{3/2}$ into a combined constant $\Gamma_2=\Gamma_{854}+\Gamma_{850}$. Thereby, one obtains the correct total decay rate of the excited state, the correct back-decay statistics, and the correct generation probability $P_{\mathrm{Raman}}$ for a Raman photon on either of the two transitions. When one is interested in the success probability of generating a Raman photon specifically at 854\,nm, this is given by $P_{854}=\Gamma_{854}/(\Gamma_{854}+\Gamma_{850})P_{\mathrm{Raman}}\approx 0.9 P_{\mathrm{Raman}}$.

\section{Decay rates of ion species}\label{appendix:rates}

The trapped-ion species considered in \autoref{sec:species} are the singly ionized alkaline-earth atoms $^{40}$Ca$^+$, $^{88}$Sr$^+$, and $^{138}$Ba$^+$, as well as the lanthanide $^{171}$Yb$^+$.  
Their relevant transition wavelengths and constants for spontaneous decays from the $\mathrm{P}_{3/2}$ and $\mathrm{P}_{1/2}$ states to the ground state and the metastable states are listed in \autoref{tab:decay} \cite{Gerritsma_2008, NIST_ASD, Gallagher_1967, Pinnington_1995, Ramm_2013, Jin_1993, Likforman_2016, Arnold_2019}.

\begin{table}[h!]\centering
\ra{1.3}
\begin{tabular}{@{}clrr@{}}\toprule
\multicolumn{2}{c}{Transition} & Wavelength (nm) & $\Gamma/2\pi$ (MHz)\\ 
\midrule
$^{40}$Ca$^+$ & $\mathrm{P}_{3/2}\to\mathrm{S}_{1/2}$ & 393 & 21.49 \\
& $\mathrm{P}_{3/2}\to\mathrm{D}_{5/2}$ & 854 & 1.35 \\
& $\mathrm{P}_{3/2}\to\mathrm{D}_{3/2}$ & 850 & 0.152\\
& $\mathrm{P}_{1/2}\to\mathrm{S}_{1/2}$ & 397 & 20.98\\
& $\mathrm{P}_{1/2}\to\mathrm{D}_{3/2}$ & 866 & 1.44\\
$^{88}$Sr$^+$ & $\mathrm{P}_{1/2}\to\mathrm{S}_{1/2}$ & 422 & 20.35\\
& $\mathrm{P}_{1/2}\to\mathrm{D}_{3/2}$ & 1092 & 1.19\\
$^{138}$Ba$^+$ & $\mathrm{P}_{1/2}\to\mathrm{S}_{1/2}$ & 493 & 14.83\\
& $\mathrm{P}_{1/2}\to\mathrm{D}_{3/2}$ & 650 & 5.43\\
$^{171}$Yb$^+$ & $\mathrm{P}_{3/2}\to\mathrm{S}_{1/2}$ & 329 & 21.74 \\
& $\mathrm{P}_{3/2}\to\mathrm{D}_{5/2}$ & 1650 & 0.24 \\
& $\mathrm{P}_{3/2}\to\mathrm{D}_{3/2}$ & 1345 & 0.037\\
\bottomrule
\end{tabular}
\caption{Transition wavelengths and spontaneous decay constants from the 
$\mathrm{P}_{3/2}$ and $\mathrm{P}_{1/2}$ state for various ion species.}
\label{tab:decay}
\end{table}

\bibliography{Bibliography.bib}
\end{document}